\documentclass{emulateapj}
\usepackage{url,amsfonts,epsfig}
\usepackage{graphicx}
\usepackage{amsmath}
\usepackage{amssymb}
\usepackage[section]{placeins}
\usepackage{chngpage}
\usepackage{calc}
\usepackage{subfigure}
\usepackage{mathtools}
\usepackage{appendix}
\usepackage{natbib}
\usepackage{graphicx}
\makeatletter 
\renewcommand\@biblabel[1]{}

\begin{document}
\def\gap{\;\rlap{\lower 2.5pt
\hbox{$\sim$}}\raise 1.5pt\hbox{$>$}\;}
\def\lap{\;\rlap{\lower 2.5pt
 \hbox{$\sim$}}\raise 1.5pt\hbox{$<$}\;}

\newcommand\sbh{MBH}
\newcommand\NSC{NSC}
\title{ROCHE-LOBE OVERFLOW IN ECCENTRIC PLANET-STAR SYSTEMS}

\author{Fani~Dosopoulou$^{1}$, Smadar Naoz$^{2,3}$, Vassiliki Kalogera$^{1}$}
\email{FaniDosopoulou2012@u.northwestern.edu}
\affil{(1) Center for Interdisciplinary Exploration and Research in Astrophysics (CIERA)
and Department of Physics and Astronomy, 
Northwestern University; \\ (2) Department of Physics and Astronomy, University of California, Los Angeles, CA 90095, USA
\\(3) Mani L. Bhaumik Institute for Theoretical Physics, Department of Physics and Astronomy, UCLA, Los Angeles, CA 90095, USA}

\begin{abstract}
Many giant exoplanets are found near their Roche limit and in mildly eccentric orbits. In this study we examine the fate of such planets through Roche-lobe overflow as a function of the physical properties of the binary components, including the eccentricity and the asynchronicity of the rotating planet. We 
use a direct three-body integrator to compute the trajectories of the lost mass in the ballistic limit and investigate the possible outcomes. We find three different outcomes for the mass transferred through the Lagrangian point $L_{1}$:  (i) self-accretion by the planet, (ii) direct impact on the stellar surface, (iii) disk formation around the star. We explore the parameter space of the three different regimes and find that at low eccentricities, $e\lesssim 0.2$, mass overflow leads to disk formation for most systems, while for higher eccentricities or retrograde orbits self-accretion is the only possible outcome. We conclude that the assumption often made in previous work that when a planet overflows its Roche lobe it is quickly disrupted and accreted by the star is not always valid.
 \end{abstract}


\section{Introduction}

Many of the shortest-period
exoplanets, covering a large mass-range, from Earth-size to Jupiter-size (hot Jupiters), are found in mildly eccentric orbits and nearly at or even interior to
their Roche limit. This suggests that many planets
are near Roche lobe overflow (RLOF) \citep[e.g., Figure 1 in][]{2017ApJ...835..145J}.
For example, \citet{2010Natur.463.1054L} suggested that WASP-12 b is in the process of RLOF  \citep[see also][]{2017arXiv170306582P}. The presence of many gas giants in similar orbits suggests that RLOF may be common among
exoplanets.

 Even if a planet is not currently close enough to its host star
to overflow, tidal interactions may eventually become important once planets are inside $\sim 0.1 \: \rm AU$ and tidal orbital decay can
drive the planet to the Roche limit
 \citep[e.g.,][]{2009ApJ...692L...9L,2009ApJ...698.1357J,2013ApJ...775L..11M,2014A&A...565L...1P}. This process tends to circularize the planet's orbit.
Furthermore, high eccentricity migration predicts that a large fraction of the planets may get disrupted. Specifically, the eccentric  Lidov-Kozai mechanism  \citep[e.g.,][]{2016ARA&A..54..441N}  can result in disrupted Jupiters \citep[e.g.,][]{2012ApJ...754L..36N,2015ApJ...799...27P}. These planets can plunge in, and cross the Roche limit, with extremely large eccentricity  or with moderate ones  that may result from planet-planet interactions \citep[e.g.,][]{2016AJ....152..174A,2016ApJ...829..132P,2017MNRAS.464..688H}.

Most previous studies have assumed that whenever a planet
crosses the Roche limit, it is quickly disintegrated and its material is accreted by the star
 \citep[e.g.,][]{2009ApJ...698.1357J,2012MNRAS.425.2778M,2013ApJ...772..143S,2014ApJ...786..139T,2014ApJ...787..131Z}. However, 
recent studies showed that hot Jupiters might be only partially consumed, leaving behind lower-mass planets \citep{2014ApJ...793L...3V,2015ApJ...813..101V, 2016CeMDA.126..227J}. These studies suggest the need for further investigation of the planetary system orbital evolution due to RLOF in eccentric systems, e.g., using secular evolution equations \citep[e.g.,][]{2016ApJ...825...70D,2016ApJ...825...71D}. The accuracy of predictions for the fate of gas giants and the final orbits of their remnants depends on the trajectory the mass lost through the Lagrangian point $L_{1}$ follows after its ejection. 
In an eccentric system the latter is determined by the eccentricity of the system as well as the masses, radii and spins of the planet-star components.
Depending on the formation history, giant exoplanets can be tidally locked or rotate asynchronously in their orbits around the star.

In this paper we study short-period eccentric planet-star systems at the onset of RLOF. We assume RLOF takes place at each subsequent periastron
passage. We investigate the possible outcomes of mass overflow
for a system with a generically asynchronous planet.
We show that for a given initial eccentricity,
depending on the binary mass-ratio, the planet rotation rate and the star radius, RLOF can lead to 
three different possible outcomes for the lost matter. These are: (i) self-accretion by the planet, 
(ii) direct impact on the stellar surface and (iii) disk formation around the star. These three different regimes do not uniquely lead to the disruption of the planet or the accretion of the lost matter 
by the star as often assumed in previous studies. Here we explore the parameter space of these regimes calculating the trajectories of the lost particle in the ballistic limit.

This paper is organized as follows. In Section 2 we describe the system and the adopted methodology. In Section 3 we explore the 
system's parameter space identifying the regions that lead to different mass overflow outcomes. We conclude with Section 4.

\begin{figure*}
\begin{center}
 \includegraphics[angle=0,width=2.3in]{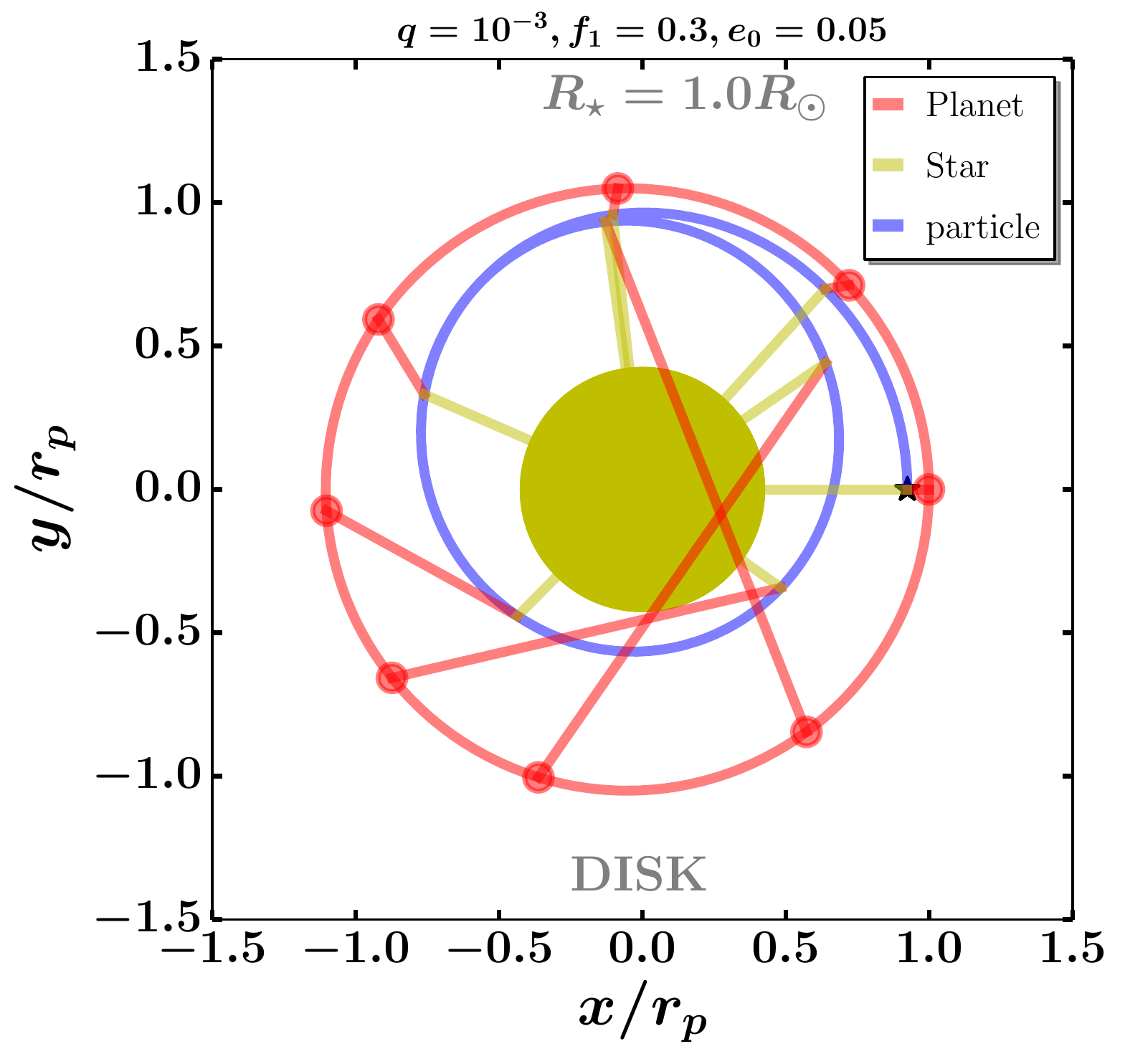}
   \includegraphics[angle=0,width=2.3in]{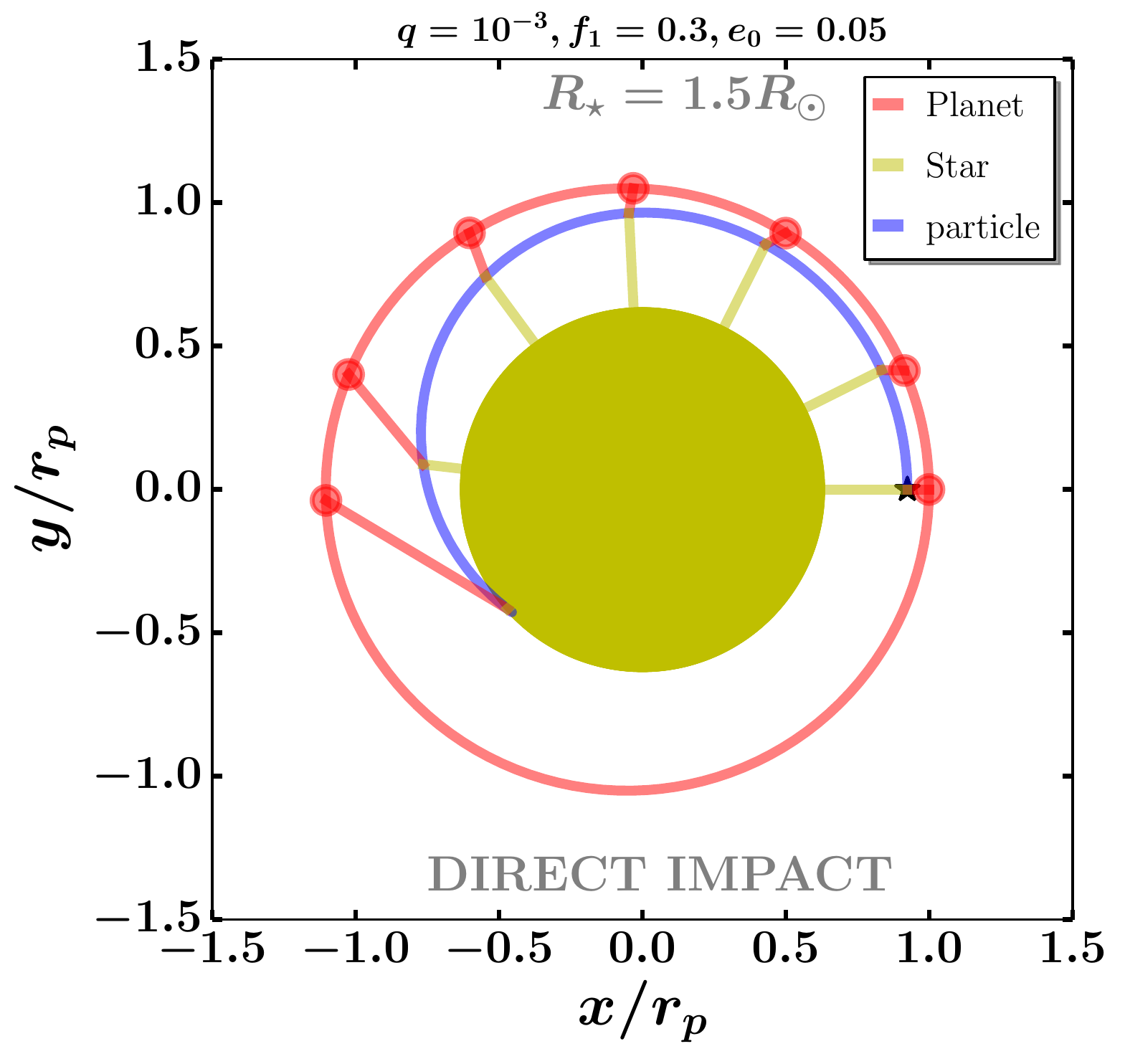}
   \includegraphics[angle=0,width=2.3in]{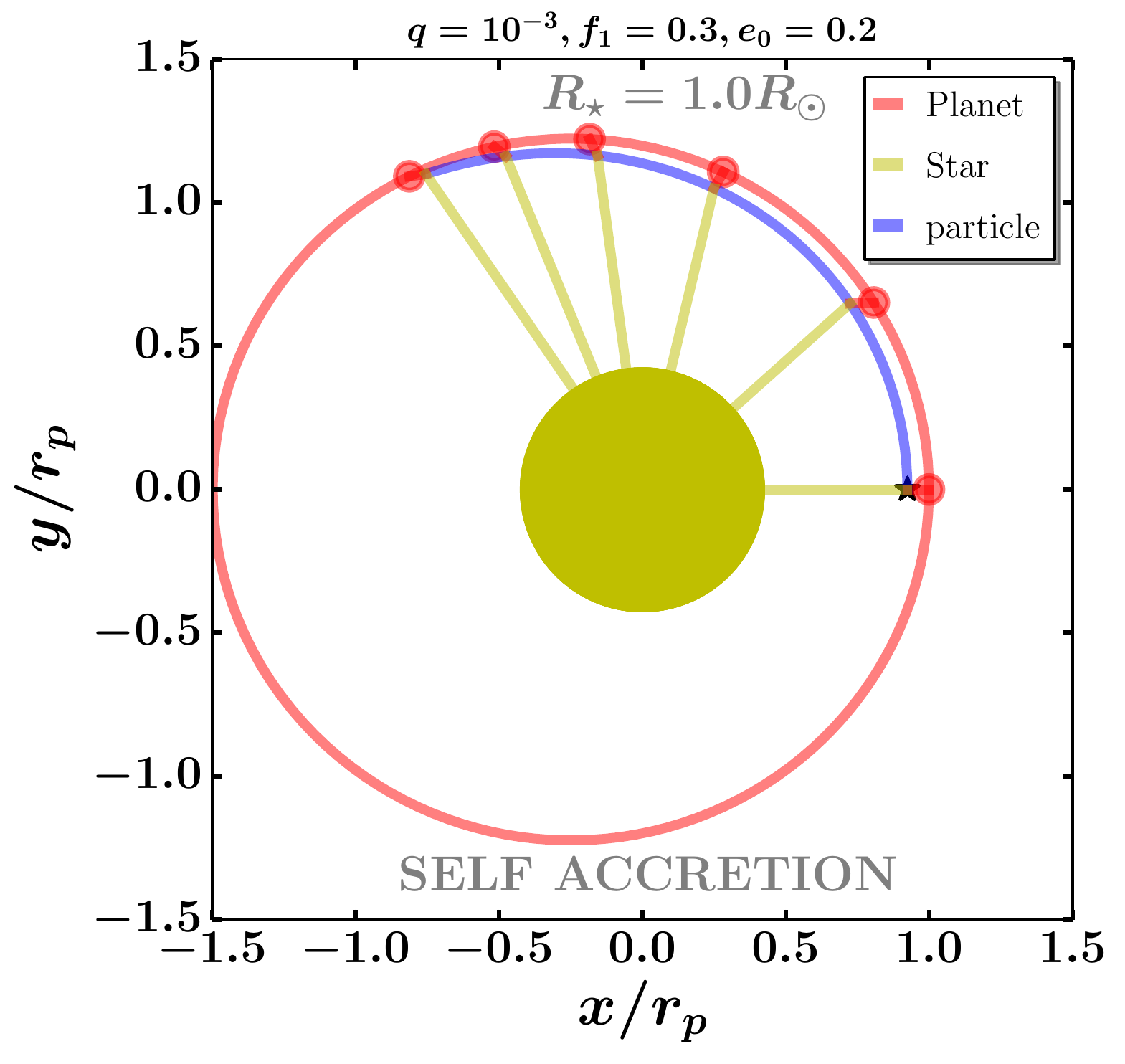}
  \caption{Different outcomes of RLOF at periastron for different initial conditions $(q,f_{\rm p},e_{0})$ and  star radii $R_{\star}=(1.0,1.5)\: R_{\odot}$ with $M_{\star}=1\: M_{\odot}$. The orbits of the planet, the star and the particle are depicted in the $(x-y)$ plane where $x$ and $y$ are normalized to the initial periastron distance $r_{\rm p}$. The location of the Lagrangian point at periastron, $x_{L_{1,\rm p}}$, is depicted with a black star while connecting lines show the relative distance of planet-particle (red) and star-particle (yellow) in different phases along the orbit. Left panel: No impact; Possibly a disk is formed (DISK). Middle panel: Direct impact (DI) on stellar surface due to the presence of a larger star. Right panel: The lost particle undergoes self-accretion (SA) due to the more eccentric orbit.}
\label{orbits}
\end{center}
\end{figure*}
\section{Roche-lobe overflow in planet-star systems}
\subsection{System set-up}
We follow the formalism developed by \citet{2010ApJ...724..546S} and consider a generically eccentric binary consisting of a planet with mass $M_{\rm p}$ and radius $R_{\rm p}$ and a star with mass $M_{\star}$ and radius $R_{\star}$. We define the binary mass-ratio as $q=M_{\rm p}/M_{\star}$ and in what follows we consider planet-star systems with mass-ratio $q$  in the range 
 $-6 \leq \log q \leq -2$. We assume that the planet rotates uniformly and with constant spin $\pmb{\Omega}_{\rm p}$ parallel or anti-parallel to the orbital angular
velocity  $\pmb{\Omega}_{\rm orb}$.
We normalize $\pmb{\Omega}_{\rm p}$ by the orbital angular velocity at periastron $\pmb{\Omega}_{\rm orb,p}$, i.e.,
\begin{equation}
\pmb{\Omega}_{\rm p}=f_{\rm p}\pmb{\Omega}_{\rm orb,p}
\end{equation}
where now $f_{\rm p}$ defines the degree of pseudo-asynchronicity of the planet at periastron and \citep[Eq. (2.32) in][]{1999ssd..book.....M}
\begin{equation}
\Omega_{\rm orb,p}=\frac{2\pi}{P_{\rm orb}}\frac{(1+e)^{1/2}}{(1-e)^{3/2}}
\end{equation}
with $P_{\rm orb}$ the binary orbital period and $e$ the binary eccentricity. For simplicity, from now on we refer to $f_{\rm p}$ as the planet's degree of asynchronicity. We consider both prograde ($f_{\rm p} >0$) and retrograde ($f_{\rm p} <0$) planetary orbits. When $f_{\rm p}=1$ the planet is pseudo-synchronously rotating at periastron, while $f_{\rm p}<1$ denotes a sub-synchronously rotating planet.

We consider short-period systems where the planet undergoes RLOF and mass is lost through the Lagrangian point $L_{1}$. \citet{2007ApJ...660.1624S} investigated the existence and properties of equipotential surfaces and Lagrangian points in asynchronous,
eccentric binary star and planetary systems. They showed that in an eccentric orbit the position of the Lagrangian point $L_{1}$ depends on the phase along the orbit and it is the smallest when the two bodies are at periastron \citep[e.g., Figure 8 in][]{2007ApJ...667.1170S}. This implies that mass overflow is more likely to reoccur at each subsequent periastron passage. In what follows we assume RLOF only at periaston and use the method developed by \citet{2007ApJ...660.1624S} to calculate the position of the Lagrangian point $L_{1}$ at periastron, $x_{L_{1,\rm p}}$. 
\citet{2007ApJ...660.1624S} provided also
fitting formulae for the volume-equivalent Roche lobe radius appropriate for asynchronous 
eccentric systems as a function of the binary mass ratio and the degree of asynchronicity of the overflowing body.
 \citet{2007ApJ...660.1624S} verified that for the low mass-ratio systems considered here, the simpler formula given in \citet{1983ApJ...268..368E} for the volume-equivalent Roche lobe radius at periastron, $R_{L_{1,\rm p}}$,  is still a good approximation for eccentric asynchronous systems \citep[e.g., Figure 9 in][]{2007ApJ...660.1624S}. Thus, we use 
\begin{equation}
R_{L_{1,\rm p}}(q)=r_{\rm p}\frac{0.49 q^{2/3}}{0.6 q^{2/3}+ \ln (1+q^{1/3})}=r_{\rm p}r_{L}(q)
\label{roche}
\end{equation}
where $r_{\rm p}=a(1-e)$ is the periastron distance and $a$ the binary semi-major axis. We assume the following planet mass $M_{\rm p}$ and radius $R_{\rm p}$ relation \citep{2011Natur.470...53L,2013Sci...340..572H}
\begin{equation}
R_{\rm p }=
\begin{cases}
R_{\rm \earth}\left(\frac{M_{\rm p}}{M_{\rm \earth}}\right)^{1/2.06} &\text{if } M_{\rm p }<M_{\rm J} ,\\\\
 R_{\rm J} &\text{if } M_{\rm p }>M_{\rm J}
\end{cases}
\label{double}
\end{equation}
where $R_{\rm \earth},M_{\rm \earth}$ are the Earth radius and mass and $R_{\rm J},M_{\rm J}$ are the Jupiter radius and mass. We also test two different mass-radius relations: (i) $R_{\rm p }= R_{\rm e}\left(M_{\rm p}/2.7M_{\rm e}\right)^{1/1.3}$  \citep{2016ApJ...825...19W}  and (ii) $R_{\rm p }=(3 M_{\rm p} / 4\pi \rho_{\rm p})^{1/3}$, where $\rho_{\rm p}$ is the planet density in the range $\rho_{\rm p}=0.1-10 \rm \:gr \:cm^{-3}$. We find that our results presented in Section 3 are not affected by the mass-radius relation adopted.

Given the initial eccentricity of the system, $e_{0}$, the initial semi-major axis, $a_{0}$,  is calculated such that the planet undergoes RLOF at periastron, i.e., for $R_{\rm p}=R_{L_{1,\rm p}}$ we find $a_{0}=R_{\rm p}/(r_{L}(q)(1-e_{0}))$. Notice that according to Equation (\ref{double}), for $M_{\rm p }>M_{\rm J}$, the initial periastron distance of the planet decreases with the mass ratio. For low initial eccentricities, $ 0.01\lesssim e_{0} \lesssim 0.2$, the value of the initial semi-major axis lies in the range $ 0.0047  \:  \textrm{AU}  \lesssim a_{0} \lesssim 0.012 \: \rm AU$.

\subsection{Ballistic limit}

We assume that a particle with negligible mass $M_{\rm loss}<< (M_{\rm p},M_{\star})$ is ejected from the planet at periastron from a point with relative distance  $x_{L_{1,\rm p}}$ to the planet center of mass and with a velocity $\pmb{V}_{\rm loss}$ relative to an inertial reference frame. 
 The planet and the star are treated as rigid spheres of uniform density and as a first approximation we can calculate the motion of the three bodies by modeling the system as three point-masses moving only under the effect of gravity. In the \emph{ballistic limit} and as long as $M_{\rm loss}<< (M_{\rm p},M_{\star})$ the  trajectories of the bodies are independent of the actual value of the lost matter. 

The ejection velocity $\pmb{V}_{\rm loss}$ of the mass transferred through the Lagrangian point $L_{1}$ is given by
\begin{equation}
\pmb{V}_{\rm loss}=\pmb{V}_{\rm p} + \pmb{\Omega}_{\rm p}\times \pmb{x}_{L_{1,\rm p}}+\pmb{V}_{\rm ej} 
\label{initialvelocity}
\end{equation}
where $\pmb{V}_{\rm p}$ is the planet orbital velocity at periastron with magnitude $V_{\rm p}\simeq \sqrt{GM_{\star}(1+e)/r_{\rm p}}$,  $ \pmb{\Omega}_{\rm p}\times \pmb{x}_{L_{1,\rm p}}$ the planet rotational velocity at $L_{1}$ and $\pmb{V}_{\rm ej} $ the ejection velocity of the
particle relative to the planet center of mass. We note here the important relation $x_{L_{1,\rm p}}>R_{L_{1,\rm p}}=R_{\rm p}$  \citep[e.g.,][]{2007ApJ...660.1624S}. This relation is key to our calculations since the ejection of the lost mass from the planet radius, $R_{\rm p}=R_{L_{1,\rm p}}$, would lead in principle the system to self-accretion (described in Section 2.3).

The ejection velocity $V_{\rm ej}$ depends on the type of mass loss. Overflow models often approximate the
donor as having a discrete outer boundary at the photosphere.  However, the upper atmospheres of close-in planets can be very hot and taper off into space. Here we make an estimate of the escape speed assuming an \emph{isothermal} atmospheric mass loss through $L_{1}$. We compare this speed to the planet's orbital velocity and show that it is reasonable to neglect the thermal speed contribution to the particle ejection velocity. The isothermal sound speed is defined
as $V_{\rm th}=\sqrt{k_{\rm B}T/\mu}$ with $k_{\rm B}$ the Boltzmann constant, $T$ the atmospheric temperature, and $\mu$ the average mass of atmospheric particles. At the planet's photosphere one finds $V_{\rm ej}<< V_{\rm th}$, while near $L_{1}$, the gas can be assumed to reach the isothermal sound speed $V_{\rm ej}\simeq V_{\rm th}$. The planet photospheric temperature $T_{\rm p}$ can be estimated by assuming that the planet dayside emits as a blackbody at radiative equilibrium with its star, i.e., $T_{\rm p}=T_{\star}\sqrt{R_{\star}/(2^{1/2}a)}$, where  $T_{\star}$ is the stellar
effective temperature. For example, for a Sun-like star with $R_{\star}=1R_{\odot}$, $M_{\star}=1M_{\odot}$ and $T_{\star} = 6000 \:\rm K$ this gives for $a\sim 0.05 \:\rm AU$ a planet temperature $T_{\rm p}=1500 \:\rm K$. For an atmosphere composed entirely of molecular hydrogen we have $\mu = 2 \rm \: amu$ when $T_{\rm p} < 2000\rm  \:\rm K$.
Using these values, the isothermal sound speed is $V_{\rm th} \sim 3500 \:\rm m/s$. The planet orbital velocity at periastron $V_{\rm p}$ assuming a small eccentricity is $V_{\rm p}\sim \sqrt {G M_{\star}/a}\sim1.3 \times 10^{5} \:\rm m/s$, i.e., $V_{\rm th} \sim 0.01 V_{\rm p}$. Note that as long as  $V_{\rm ej}/ V_{\rm p} \lesssim 0.01$, the angular momentum
exchange between the particle and the binary during transport is unaffected by changes in $V_{\rm ej}$. Given these considerations, here we assume no contribution to the particle ejection velocity from the thermal speed of the mass elements in the planet atmosphere and set $V_{\rm ej}=0$.

We note that atmospheric mass loss is neither entirely adiabatic nor isothermal. In principle there is also a transition between RLOF and evaporative mass loss. In the latter, the temperature in the upper atmosphere can greatly exceed $\sim 10^{4}\:\rm K$  and the outflow becomes transonic \citep{2009ApJ...693...23M,2011ApJ...728..152T}. However, recent studies have shown that for the vast majority of the systems considered here, if atmospheres are escaping at all, it is via RLOF \citep{2017ApJ...835..145J}. A generalization of our calculations would account for the thermal speed in the planet atmosphere but this will add an extra dimension to the parameter space and it is beyond the scope of this paper.

In what follows we use a direct three-body code (developed by \citet{2010ApJ...724..546S}) to integrate the equations of motion forward in time and calculate the ballistic trajectories of the lost particle.

\subsection{Possible outcomes of Roche-lobe overflow}

We evolve the planet-star-particle system for one orbit. We keep track of the particle distance to the planet and the star as a function of time and, as depicted in Figure \ref{orbits}, we find three possible outcomes for the particle lost through $L_{1}$ at periastron. These are: (i) the lost particle is self-accreted by the planet within one orbit (SA), (ii) the lost particle directly impacts the stellar surface within one orbit (DI), (iii) the lost particle undergoes no impact with the planet or the star and its ballistic trajectory intersects itself within one orbit leading possibly to the formation of a disk around the star through interactions with subsequently lost particles (DISK).

In Figure \ref{orbits} we show characteristic examples of the three aforementioned mass overflow outcomes for different initial conditions $(q,f_{\rm p},e_{0})$ and  for star radii $R_{\star}=(1.0,1.5)\: R_{\odot}$ and mass $M_{\star}=1 M_{\odot}$. As shown in Figure \ref{orbits} in the two systems in the left and middle panel the lost particle follows the same ballistic trajectory.
However, in these systems a star with a radius $R_{\star}=1\: R_{\odot}$ leads to the formation of a DISK while  a star with  a larger radius $R_{\star}=1.5R_{\odot}$ results in DI. In an initially more eccentric system, the lost particle has a larger ejection velocity which leads to SA.

We also explore retrograde systems for which $f_{\rm p}<0$. In this case we find that, as expected from Equation (\ref{initialvelocity}), the particle is always moving faster than the planet at the point of ejection.  This leads always to SA independent of the eccentricity. In what follows we focus on prograde orbits ($f_{\rm p}>0$).

\section{Results}
\begin{figure*}
    \includegraphics[width=1.0\textwidth]{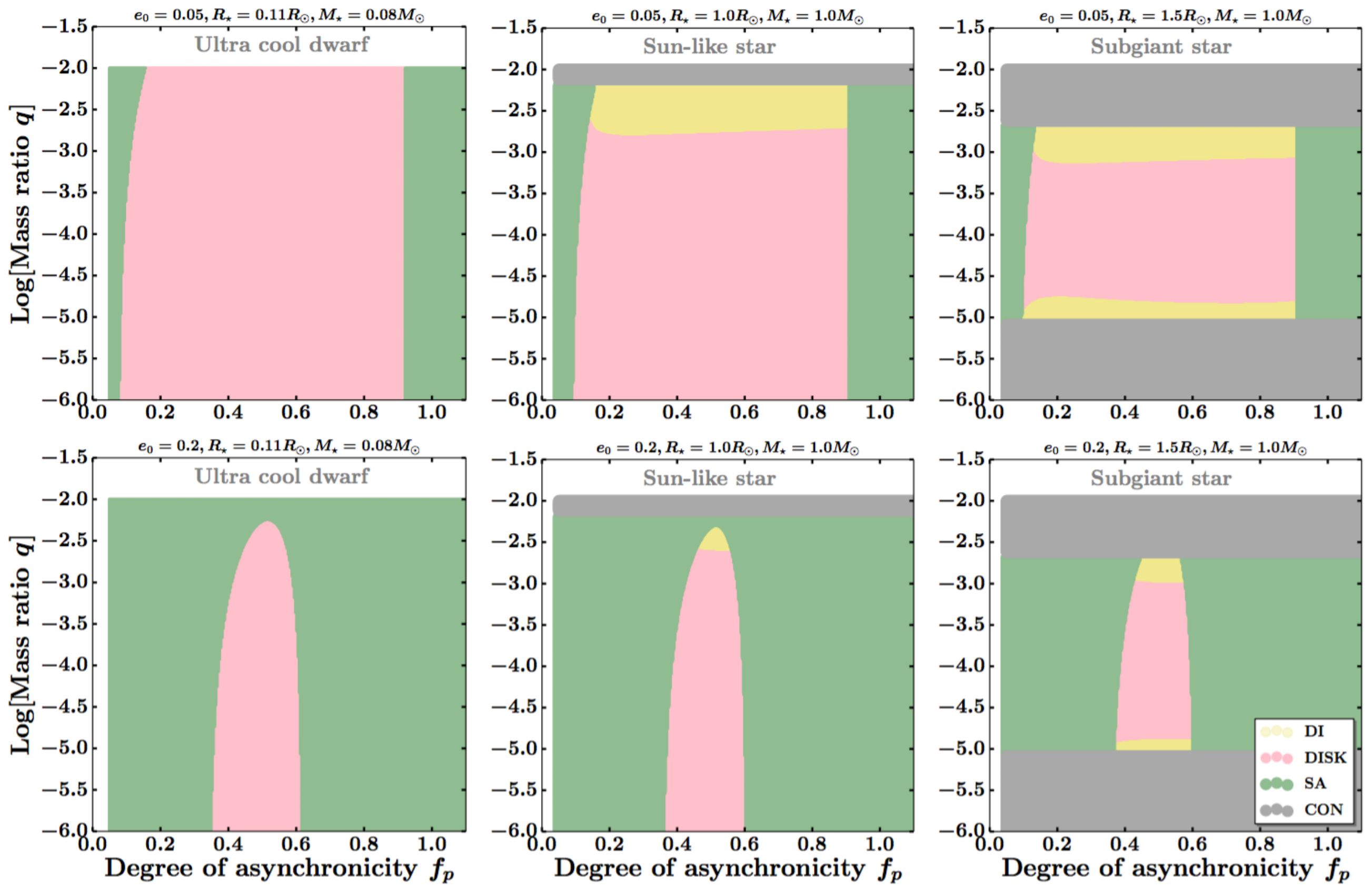}
      \caption{RLOF at periastron outcomes for $e_{0}=0.05$ (top panels), $e_{0}=0.2$ (bottom panels) and a ultra cool dwarf star with $M=0.08 M_{\odot}$ and $R_{\star}=0.11 R_{\odot}$ (left column) a Sun-like star with mass $M=1 M_{\odot}$ and radius  $R_{\star}=1.0 R_{\odot}$ (middle column) a subgiant star with  $M=1 M_{\odot}$ and  $R_{\star}=1.5  R_{\odot}$ (right column). Grey areas indicate systems where the planets comes into contact with the star (CON). DISK/DI occurs mostly at low eccentricities while at higher eccentricities SA takes over. For the types of systems considered here, although an increase in the stellar radius leads to an extension of the DI and CON regime, DISK formation dominates over DI for both eccentricities.}
      \label{types}
\end{figure*}

In this section we explore the system parameter space and investigate the dependence of the mass overflow outcome on the initial conditions $(q,f_{\rm p},e_{0})$  as well as the mass $M_{\star}$ and radius $R_{\star}$ of the star.
 
In Figure \ref{types} we explore the mass ratio and asynchronicity parameter space of the system  $(q,f_{\rm p})$ for two different initial eccentricities $e_{0}$ and three different types of stars: (i) an ultra-cool dwarf star with $M_{\star}=0.08 M_{\odot}$ and radius  $R_{\star}=0.11 R_{\odot}$ similar to the one in the recently discovered TRAPPIST-1 planetary system \citep{2017Natur.542..456G} (ii) a Sun-like star with mass $M_{\star}=1 M_{\odot}$ and radius  $R_{\star}=1.0 R_{\odot}$ (iii) a subgiant star with mass $M_{\star}=1 M_{\odot}$ and radius $R_{\star}=1.5 R_{\odot}$.  

As shown in Figure \ref{types} for a given eccentricity, $e_0$, the values of $f_{\rm p}$ 
and $q$  which separate the SA and DISK/DI regimes do not depend on the mass $M_{\star}$ and radius $R_{\star}$ of the star. At low eccentricities, the most
likely outcome is DISK/DI, while at higher eccentricities, $e\gtrsim 0.2$, SA takes over. Although increasing the star radius, $R_{\star}$, increases the parameter space 
for DI, for the cases we consider in Figure \ref{types}, the formation of a DISK appears to be a more common outcome than DI.

For the systems we consider in Figure \ref{types}, at a given $R_{\star}$, there are values of  $q$ such that $R_{\star} > r_{\rm p}-x_{L_{1,\rm p}}$, i.e., the planet comes into contact with the star. The parameter space for these type of systems is shown in Figure \ref{types} as a grey area (CON).
For planets less massive than Jupiter, the periastron distance increases with a decreasing mass ratio. This implies that for a small enough mass ratio the planet comes into contact with the stellar surface. This is depicted as the lower gray area in the right column in Figure \ref{types}, where for the subgiant star we estimate this lower limit to be $\log q \sim -5.25$. For planets more massive than Jupiter, we adopted an upper limit for the planet radius  ($R_{\rm p }=R_{\rm J}$). This means that for these planets the periastron distance decreases with an increasing mass ratio. Thus, for large mass ratio above a threshold the planet and the star come into contact. This is depicted as the upper gray area in the right column in Figure \ref{types} where for the subgiant star this upper threshold becomes $\log q \sim -2.6$.

\begin{figure}
    \includegraphics[width=0.45\textwidth]{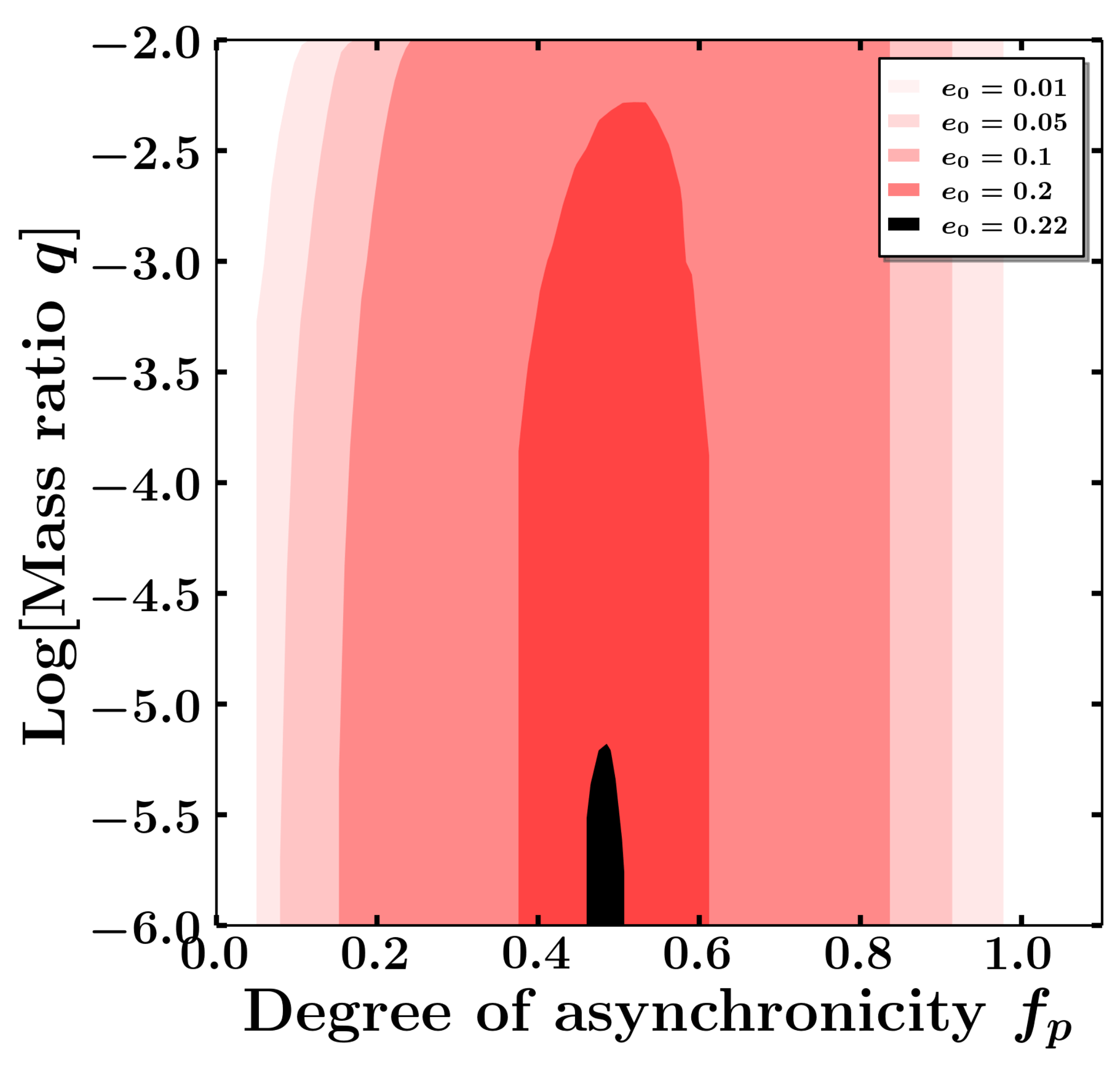}
      \caption{Color-shaded regions depict the parameter space $(q,f_{\rm p})$ for which either DI or DISK occurs.  Outside these areas SA takes over. Darker tone of red refers to larger initial eccentricity which restricts the DI/DISK regime. DI/DISK dominates at low eccentricities  while for higher eccentricities, $e_{0} \gtrsim 0.2$, the only possible outcome is SA. Black region refers to the highest eccentricity that allows for DISK/DI.}
      \label{spaceshade}
    \includegraphics[width=0.45\textwidth]{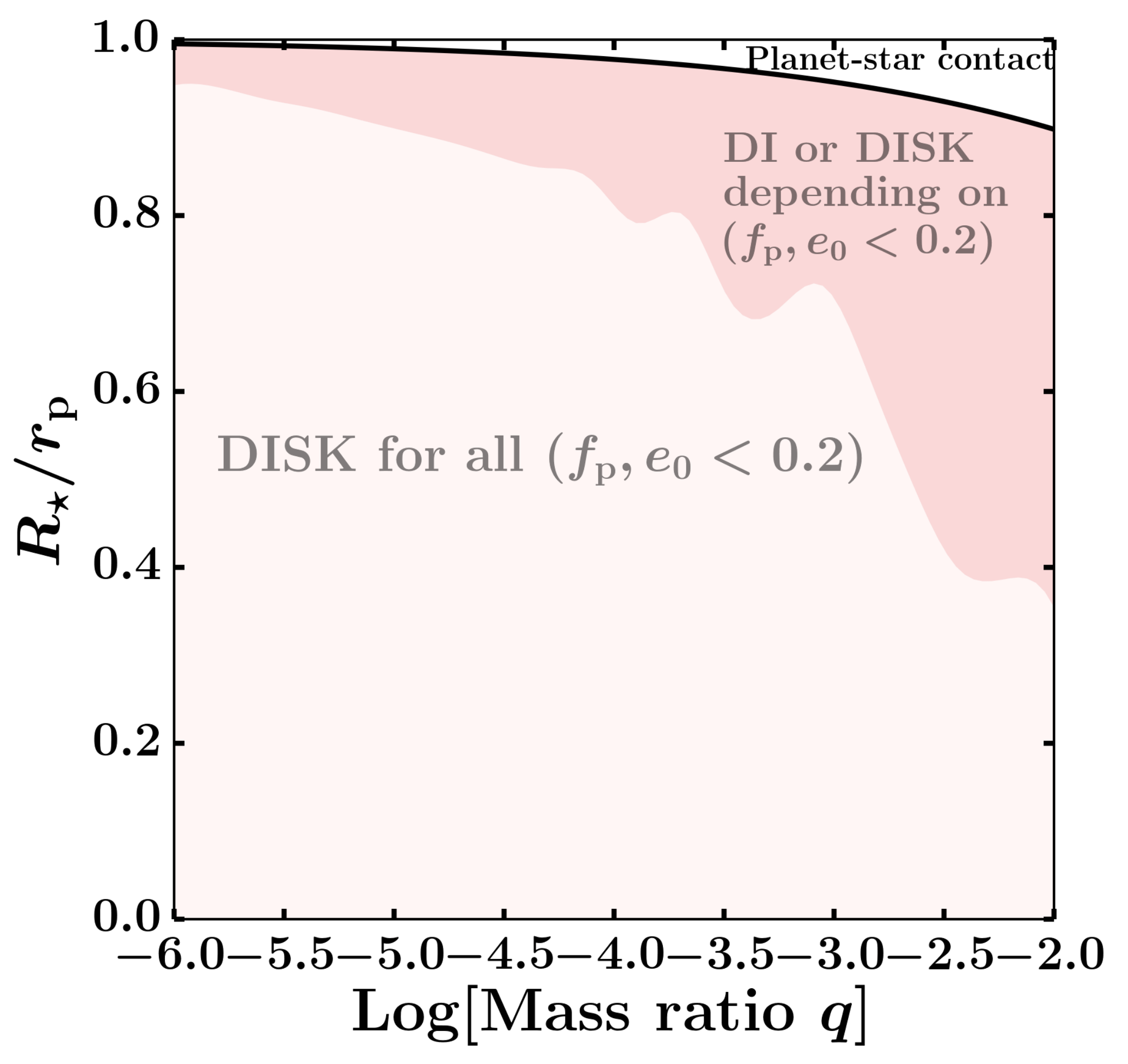}
      \caption{Here we consider low-eccentricity systems, $e_{0}<0.2$, and set  $M_{\star}=1 M_{\odot}$. Color-shaded regions separate the ``DISK" regime (always a DISK forms regardless of $(f_{\rm p},e_{0})$) from the ``DI or DISK" regime (DISK or DI depending on $(f_{\rm p},e_{0})$). The black line refers to $R_{\star,\rm max}(q)$, i.e., in the white area the planet and the star are in contact. At low eccentricities RLOF at periastron leads to DISK formation for most systems.}
      \label{didisk}
\end{figure}

In Figure \ref{spaceshade} we investigate the parameter space that separate the SA and DI/DISK regimes.
As depicted in Figure \ref{types} the values of $f_{\rm p}$ and $q$  at which the transition to SA occurs do not depend on $M_{\star}$ 
and $R_{\star}$. Thus, for simplicity, we set in Figure \ref{spaceshade} $R_{\star}=0$ and $M_{\star}=1 M_{\odot}$, and explore the $(q,f_{\rm p})$ parameter space  as a function of the  initial eccentricity, $e_0$.
As Figure \ref{spaceshade} indicates, for any initial eccentricity, DI/DISK can occur only for sub-synchronously rotating planets ($f_{\rm p}< 1$). Increasing the initial eccentricity restricts the parameter space region for DI/DISK. We find that the DI/DISK regime dominates at low eccentricities, $e_{0} \lesssim 0.2$, while for higher eccentricities or retrograde orbits the only possible outcome is SA. 

As shown in Figure \ref{spaceshade}, DI/DISK can occur only for $e_{0}<0.2$. 
Within this restricted regime we identify in Figure \ref{didisk}  two regions, depending on the star radius $R_{\star}$ and the mass-ratio $q$. As we mentioned before, for a given $q$, there exists an upper limit to the radius of the star, $R_{\star,\rm max}$, above which the planet and the star are in contact. Here we compute this upper limit setting $R_{\star,\rm max}=r_{\rm p}-R_{L_{1,\rm p}}$ (notice that for low mass ratios $x_{L_{1,\rm p}}\simeq R_{L_{1,\rm p}}$). This upper limit $R_{\star,\rm max}$ is plotted in Figure  \ref{didisk} as a function of the mass ratio $q$. If the star radius $R_{\star}$ falls within the region labeled as ``DISK",  mass overflow leads always to the formation of a disk regardless of the values chosen for $f_{\rm p}$ and $e_{0}$. This is because within the ``DISK" regime the star radius $R_{\star}$ is always smaller than the minimum particle distance to the star.
In the region labeled as ``DI or DISK" the mass overflow outcome can be either DISK or DI depending  on the specific values of $f_{\rm p}$ and $e_{0}$. As depicted in Figure  \ref{didisk}, as a star begins to expand without losing mass (e.g., as the star leaves the Main Sequence), a planet undergoing RLOF in a low-eccentricity orbit may lead to DI, before the planet gets in contact with the stellar surface. However, as shown in Figure \ref{didisk}, the region for which a DISK forms overall dominates the parameter space. This implies that at low eccentricities, RLOF at periastron leads to DISK formation for most systems.

\section{Conclusions}

We investigated the possible outcomes of Roche lobe overflow  at periastron for eccentric planet-star systems with a planet in an asynchronous orbit. We explore the system parameter space identifying the regimes that lead to different outcomes of the planet's mass loss.\\

The main results of this paper are summarized below:

\begin{itemize}

\item[1)] Roche lobe overflow at periastron leads to one of the three possible outcomes for the mass transferred though the Lagrangian point $L_{1}$: (i) self-accretion by the planet (ii) direct impact on the stellar surface  (iii) disk formation around the star (see Figure \ref{orbits}).

\item[2)] Direct impact or disk formation can occur only at low eccentricities and only for systems with sub-synchronously rotating planets in prograde orbits. For higher eccentricities, $e\gtrsim 0.2$,  mass overflow leads always to self-accretion (see Figure \ref{spaceshade}). In the case of retrograde orbits the only possible outcome is SA independent of the system eccentricity.

\item[3)] For the low eccentric planet-star systems, $e\lesssim 0.2$, and within the direct impact/disk formation parameter space regime, the region where a disk is formed dominates the region that leads to direct impact (see Figure \ref{types}).  Although increasing the star radius for a given stellar mass (e.g., as the star evolves beyond the Main Sequence) may lead to direct impact, at low eccentricities Roche lobe overflow leads to disk formation for most systems (see Figure \ref{didisk}).
\end{itemize}
We have considered as a proof of concept the ballistic approach to study Roche-lobe overflow in eccentric planet-star systems. We have shown that at low eccentricities Roche lobe overflow leads to disk formation for most systems. For eccentric systems we speculate that the dynamics may lead to the survival of giant planets near the Roche limit, as observed \citep{2017ApJ...835..145J}. Using the formalism described in \citet{2016ApJ...825...70D,2016ApJ...825...71D} the secular evolution of these systems can be investigated to test the speculation mentioned above and to compare to observations.

\section*{ACKNOWLEDGEMENTS}
We want to thank our colleague Jeremy Sepinsky who shared with us the three-body code and Fred Rasio for useful discussions.
F.D. and V.K. acknowledge support from grant NSF AST-1517753 and Northwestern University through the “Reach for
the Stars” program. S.N. acknowledges partial support from a Sloan Foundation Fellowship.

\end{document}